\documentclass[twoside,twocolumn,9pt]{article}
\usepackage{extsizes}
\usepackage[version=3]{mhchem}
\usepackage[left=1.5cm, right=1.5cm, top=1.785cm, bottom=2.0cm]{geometry}
\usepackage{balance}
\usepackage{times,mathptmx}
\usepackage{sectsty}
\usepackage{graphicx} 
\usepackage{lastpage}
\usepackage[format=plain,justification=justified,singlelinecheck=false,font={stretch=1.125,small,sf},labelfont=bf,labelsep=space]{caption}
\usepackage{float}
\usepackage{fancyhdr}
\usepackage{fnpos}
\usepackage[english]{babel}
\usepackage{array}
\usepackage{droidsans}
\usepackage{charter}
\usepackage[T1]{fontenc}
\usepackage[usenames,dvipsnames]{xcolor}
\usepackage{setspace}
\usepackage[compact]{titlesec}

\usepackage{epstopdf}

\definecolor{cream}{RGB}{222,217,201}

\usepackage{mathrsfs}
\usepackage{amsmath}
\usepackage{amssymb}
\usepackage[usetitle]{rsc}
\usepackage{datetime}

\newcommand{\beq}{\begin{eqnarray}}
\newcommand{\eeq}{\end{eqnarray}}
\newcommand{\dd}{{ \! \! \rm d}}

\newcommand{\un}[1]{{\rm #1}} 
\newcommand{\tw}{\columnwidth}

\begin{document}

\pagestyle{fancy}
\thispagestyle{plain}
\fancypagestyle{plain}{
\renewcommand{\headrulewidth}{0pt}
}

\makeFNbottom
\makeatletter
\renewcommand\LARGE{\@setfontsize\LARGE{15pt}{17}}
\renewcommand\Large{\@setfontsize\Large{12pt}{14}}
\renewcommand\large{\@setfontsize\large{10pt}{12}}
\renewcommand\footnotesize{\@setfontsize\footnotesize{7pt}{10}}
\makeatother

\renewcommand{\thefootnote}{\fnsymbol{footnote}}
\renewcommand\footnoterule{\vspace*{1pt}%
\color{cream}\hrule width 3.5in height 0.4pt \color{black}\vspace*{5pt}} 
\setcounter{secnumdepth}{5}

\makeatletter 
\renewcommand\@biblabel[1]{#1}            
\renewcommand\@makefntext[1]%
{\noindent\makebox[0pt][r]{\@thefnmark\,}#1}
\makeatother 
\renewcommand{\figurename}{\small{Fig.}~}
\sectionfont{\sffamily\Large}
\subsectionfont{\normalsize}
\subsubsectionfont{\bf}
\setstretch{1.125} 
\setlength{\skip\footins}{0.8cm}
\setlength{\footnotesep}{0.25cm}
\setlength{\jot}{10pt}
\titlespacing*{\section}{0pt}{4pt}{4pt}
\titlespacing*{\subsection}{0pt}{15pt}{1pt}

\fancyfoot{}
\fancyfoot[RO]{\footnotesize{\sffamily{1--\pageref{LastPage} ~\textbar  \hspace{2pt}\thepage}}}
\fancyfoot[LE]{\footnotesize{\sffamily{\thepage~\textbar\hspace{0.01cm} 1--\pageref{LastPage}}}}
\fancyhead{}
\renewcommand{\headrulewidth}{0pt} 
\renewcommand{\footrulewidth}{0pt}
\setlength{\arrayrulewidth}{1pt}
\setlength{\columnsep}{6.5mm}
\setlength\bibsep{1pt}

\makeatletter 
\newlength{\figrulesep} 
\setlength{\figrulesep}{0.5\textfloatsep} 

\newcommand{\topfigrule}{\vspace*{-1pt}%
\noindent{\color{cream}\rule[-\figrulesep]{\columnwidth}{1.5pt}} }

\newcommand{\botfigrule}{\vspace*{-2pt}%
\noindent{\color{cream}\rule[\figrulesep]{\columnwidth}{1.5pt}} }

\newcommand{\dblfigrule}{\vspace*{-1pt}%
\noindent{\color{cream}\rule[-\figrulesep]{\textwidth}{1.5pt}} }

\makeatother

\twocolumn[
  \begin{@twocolumnfalse}
\vspace{3cm}
\sffamily
\begin{tabular}{m{4.5cm} p{13.5cm} }
& \noindent\LARGE{\textbf{Continuous and discontinuous morphological transitions between capillary bridges on a beaded chain pulled out from a liquid.$^\dag$}} \\
\vspace{0.3cm} & \vspace{0.3cm} \\
& \noindent\large{Filip Dutka,\textit{$^{a, \, b}$} Zbigniew Rozynek,\textit{$^{c}$} and Marek Napi\'orkowski\textit{\,$^{a}$}} \\
& \noindent\normalsize{We describe theoretically and validate experimentally the mechanism of formation of capillary bridges during pulling a beaded chain out from a liquid with a planar surface. There are two types of capillary bridges present in this system, namely the sphere-planar liquid surface bridge initially formed between the spherical bead leaving the liquid bath and the original bulk planar liquid surface, and the sphere-sphere capillary bridge formed between neighbouring beads in the part of the chain above the liquid surface. During pulling the chain out of the liquid, the sphere-planar liquid surface bridge transforms into the sphere-sphere bridge. We show that for monodisperse spherical beads comprising the chain, this morphological phase transition can be either continuous or discontinuous. The transition is continuous when the diameter of the spherical beads is larger than the capillary length. Otherwise, the transition is discontinuous, likewise the capillary force acting on the chain. } 
\end{tabular}

 \end{@twocolumnfalse} \vspace{0.6cm}

  ]

\renewcommand*\rmdefault{bch}\normalfont\upshape
\rmfamily
\section*{}
\vspace{-1cm}


\footnotetext{\textit{$^{a}$~Institute of Theoretical Physics, Faculty of Physics, University of Warsaw, Pasteura 5, 02-093 Warszawa, Poland. Tel: +48 2255 32908; E-mail: fdutka@fuw.edu.pl}}
\footnotetext{\textit{$^{b}$~Institute of Physical Chemistry, Polish Academy of Sciences, Kasprzaka 44/52, 01-224 Warszawa, Poland. }}
\footnotetext{\textit{$^{c}$~Faculty of Physics, Institute of Acoustics, Adam Mickiewicz University, Umultowska 85, 61-614 Pozna\'n, Poland. }}

\footnotetext{\dag~Electronic Supplementary Information (ESI) available: one movie and two animations.}



\section{Introduction}

Capillary bridges play important role in many physical phenomena, including agglomeration \cite{Balakin2015,Wang2015,Balakin2013}, mechanical strengthening \cite{Herminghaus2005,Sheel2008,Pakpour2012}, surface adhering \cite{Xue2015,Slater2014}, rheological response \cite{Koos2011,Koos2012,Hoffmann2014}, capillary-gripping \cite{Fan2015,Vasudev2008} and self-assembly \cite{Arutinov2014,Broesch2014,Mastrangeli2015}. They are often utilized in material science, e.g. for fabricating new materials using capillary suspensions \cite{Velankar2015,Dittmann2015,Schneider2016}, new structures \cite{Park2012,Wang2015b,Tisserant2015}, in nanolithography \cite{Chaix2006,Fabie2010,Eichelsdoerfer2014}, microplating \cite{Hunyh2013}, for pattern formation \cite{Xu2006}, and in printed electronics technologies \cite{Kumar2015}. It is thus essential to know their properties and behavior. Therefore, a great effort is made to understand the mechanisms of their formation and shape development \cite{Anachkov2016,Wu2016,Gogelein2010}, rupturing \cite{Perales2011,Alexandrou2010,Yang2010,Men2011} and evaporation \cite{Cho2016,Neeson2014}. 

Capillary bridges exist at solid contacts between spheres \cite{Bayramli1987,Adams2002,Lian2016}, rods \cite{Mollot1993,Duprat2012,Sauret2015}, plates \cite{Dejam2015,Cheng2016}, a mix of these \cite{Rabinovich2005,Dutka2007,Guzowski2010,Dormann2014}, or other shapes \cite{Wang2016,Dutka2006}. 
They are also formed, when a particle is being pulled out from a suspension. Then a liquid rises to a certain height above the bulk planar level, and forms a concave meniscus. This meniscus is called a capillary bridge between a particle and a planar liquid surface. Such capillary bridges can be used for determining the surface tension \cite{Huh1976,Bayramli1982,He2015,Ettelaie2015}, in a way similar to the methods that employ pulling the so-called Wilhelmy plate \cite{Butt2006,Butt2009b,Jorgensen2015}, a cylindrical fiber \cite{Takahashi1990,Quere1999,Gennes2004}, or a toroidal ring \cite{Hubbard2002,Drelich2002}.

\begin{figure}[htb]
 \begin{center}
  \includegraphics[width = \tw]{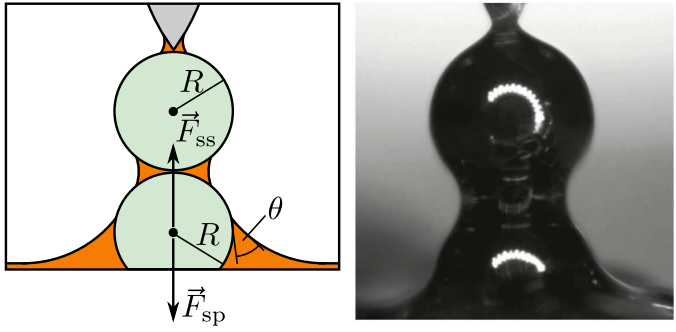}
	\caption{Scheme illustrating the capillary forces acting on the lower spherical particle of radius $R$: sphere-planar liquid surface capillary force directed downwards $\vec{F}_\un{sp}$, and sphere-sphere capillary force directed upwards $\vec{F}_\un{ss}$. We assume the contact angle $\theta=0$, i.e. the bridge is tangential to the sphere (left panel). A magnified image of two spheres aligned along the direction of pulling, which is upwards. The upper particle is the beginning of a chain. The lower particle is just being pulled out from liquid. The liquid layer forming a capillary sphere-sphere bridge is well-resolved (right panel). \label{fig_scheme}}
 \end{center}
\end{figure}

In this article, we study the formation of a solid-planar liquid surface bridge and its morphological transition into a solid-solid liquid capillary bridge. We analyze the development of these capillary bridges formed on a beaded chain being pulled out from a liquid, Fig\,\ref{fig_scheme}. This research originates from the observation of the formation of liquid bridges during a novel process of fabricating one-dimensional colloidal assemblies in the presence of dipolar interactions \cite{Rozynek2016}. We observed that the assembly process proceeded unevenly when the beads forming the chain were small (micrometers) and smoothly when the beads had sub-millimeter size, see Supplementary Movie 1. The way this process proceeded was related to the mechanism of the capillary bridges formation. Here we provide the theoretical description of that mechanism, and we show that for monodispersed spherical beads forming a chain, the morphological phase transition between two types of bridges can be either continuous or discontinuous. The transition is continuous when the diameter of the spheres is larger than the capillary length $\lambda=\sqrt{\gamma/\rho_\un{l} g}$, where $\gamma$ is surface tension coefficient, $\rho_\un{l}$ liquid  density, and $g$ gravitational acceleration. Otherwise, the transition is discontinuous, as is the capillary force acting on the chain.

In order to experimentally validate our theoretical predictions we prepared experiments with two beaded chains composed of either large ($2 \, \un{mm}$) or small ($30 \, \un{\mu m}$) spheres (capillary length $\lambda=1.45 \, \un{mm}$). We either glued the spheres to form a permanent chain, or assembled particle by employing the process described in reference \cite{Rozynek2016}. In short, the process requires the sum of attractive dipolar force and capillary forces between neighbouring particles $F_\un{ss}$ to be sufficiently strong to overcome the capillary force $F_\un{sp}$ stemming from a sphere-planar liquid surface bridge. Capillary bridges stabilize the growing chain.

This article is structured as follows: in Sec.\,\ref{sec_sphereplane} we introduce the theoretical model that is then used to describe the formation of the sphere-planar liquid surface bridge, calculate the shape of the bridge in the presence of gravitational field, and the corresponding capillary force. We track the morphological phase transition to the sphere-sphere bridge phase. The transition can be continuous or discontinuous, depending on the size of radius of spheres and the capillary length, which is discussed in Sec.\,\ref{sec_transitions}. Both the sphere-liquid surface and sphere-sphere capillary forces are  evaluated and discussed in Sec.\,\ref{sec_forces}. In Sec.\,\ref{sec_experiment}, we compare our theoretical predictions with the experiment, and close the paper with a short summary in Sec.\,\ref{sec_summary}.


\section{Liquid bridge between a sphere and a planar liquid surface \label{sec_sphereplane}}

\begin{figure}[htb]
 \begin{center}
  \includegraphics[width = \tw]{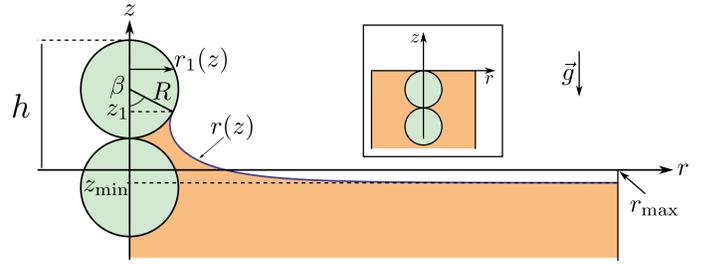}
	\caption{Schematic shape of a sphere-planar liquid surface bridge formation, during pulling out a beaded chain from a liquid bath of size $r_\un{max}$, and initial flat configuration of the liquid surface (inset). The chain consists of aligned in line, touching each other spherical beads of radius $R$. The shape of the bridge is cylindrically symmetric and described by a function $r(z)$. The maximum level of the liquid meniscus is $z_1$, and the minimum is $z_\un{min}$. Upon increasing the height $h$, the characteristic angle $\beta$ decreases. The system is placed in a gravitational field $\vec{g}$. \label{fig_sp2}}
 \end{center}
\end{figure}

The initial level of the planar liquid surface is assumed to be at $z=0$, see inset in Fig.\,\ref{fig_sp2}. The chain is composed of spheres of radius $R$ that touch one another, and are aligned in line along the $z$-axis. The system is assumed to have cylindrical symmetry around the chain axis, and the radius of the container is $r_\un{max}$. Upon pulling out the chain, the sphere-planar liquid surface bridge emerges, and, because the volumes of both the liquid and the spheres are fixed, the  liquid level decreases from $z=0$ to $z=z_\un{min}<0$. 
The energy of the sphere-planar liquid surface bridge, as compared to the initial planar configuration of the liquid surface, is a sum of capillary and gravitational terms
\begin{align} \label{eq_functional}
 \begin{split}
E_\un{sp}[r(z)] &= 2 \pi \int_{z_\un{min}}^{z_1} \dd z \Bigg[ \gamma \, r(z) \sqrt{1+r'(z)^2} \\
	 	 &  \hspace{2.5cm} +  \frac{1}{2} \rho_\un{l} g \, z \Big(r(z)^2-r_1(z)^2 \Big)\Bigg]  \\
		& -  \pi r_\un{max}^2 \gamma + 2 \pi  R (h-z_1) (\gamma_\un{sg}-\gamma_\un{sl})  \, ,
\end{split}
\end{align}
with the constant liquid volume constraint 
\begin{align} \label{eq_constraint}
 \begin{split}
  \Delta V[r(z)] = & \pi \int_{z_\un{min}}^{z_1} \dd z \, r(z)^2 + \frac{\pi}{3}(h-z_1)^2(3R-h+z_1) \\
	&  - \pi r_\un{max}^2 |z_\un{min}| = 0 \, .
\end{split}
\end{align}
Parameters $\gamma$, $\gamma_\un{sg}$, $\gamma_\un{sl}$ are the liquid-gas, sphere-gas, and sphere-liquid surface tension coefficients, respectively. The density of the liquid is $\rho_\un{l}$, $g = 9.81 \, \un{m/s^2}$ -- gravitational constant, and $r=r_1(z)$ describes the surface of  the spheres already pulled out above the planar liquid surface. The quantity $\Delta V$ is the difference of volumes corresponding to the sphere-planar liquid surface bridge configuration, and the one corresponding to the initial flat surface configuration. Note that in both configurations, the volume of the spheres is taken into account.

We assume the contact angle $\theta=0$, i.e. the bridge meniscus is tangential to the sphere at $z=z_1$. This approximates experimental situations in which one observes very small values of the contact angle. From the Young's equation, one has
\beq
 \gamma_\un{sg}-\gamma_\un{sl} = \gamma \cos \theta = \gamma \, ,
\eeq    
and the last term in Eq.\,(\ref{eq_functional}) describing the energy of the sphere covered with the thin liquid film  ($z>z_1$) reduces to $2 \pi  R (h-z_1) \gamma$.

For a given height $h$, the equilibrium profile $r(z)=r_\un{eq}(z)$ minimizes the functional $E_\un{sp}[r(z)]$, Eq.\,(\ref{eq_functional}), under the constant volume constraint
\beq 
 0 = \left. \frac{\delta  \Big(E_\un{sp}[r(z)] - \Delta p \, \Delta V[r(z)] \Big)}{\delta r(z)} \right|_{r(z)=r_\un{eq}(z)}\, ,
\eeq
which gives the equation for the shape of the interface
\begin{align} \label{eq_shape}
 \frac{1}{r_\un{eq} (1+r'_\un{eq}(z)^2)^{1/2}} - \frac{r''_\un{eq}(z)}{(1+r'_\un{eq}(z)^2)^{3/2}} = \frac{\Delta p}{\gamma} - \frac{z}{\lambda^2} \, .
\end{align}
The expression on the left hand side of the above equation represents the mean curvature multiplied by factor two, $\Delta p$ is the Lagrange multiplier, which is the difference of pressures of the inner liquid phase and the outer gas phase (Laplace pressure). If the hydrostatic pressure corresponding to height $z$, i.e., $z \, \rho_\un{l} \, g$ is small compared to the Laplace pressure $\Delta p$, then the liquid meniscus forms a constant mean curvature surface \cite{Langbein2002,Boucher1980}, and can be described analytically by elliptic integrals \cite{Kralchevsky2001,Honschoten2010}. In our analysis we take into account the the presence of gravitational field, and the shape of the bridge can be determined only numerically.


\subsection{Equilibrium and metastable states}

For large containers, $r_\un{max} \gg \lambda, R$, one can consider the limiting case of infinite system, $r_\un{max} \to \infty$. For such a case, there is no fixed volume constraint, thus, the pressure difference vanishes, $\Delta p =0$. 
The equilibrium shape of the interface  $r_{0}(z)$ fulfills the equation 

\begin{align} \label{eq_shape0}
 \frac{1}{r_0 (1+r_0'(z)^2)^{1/2}} - \frac{r_0''(z)}{(1+r_0'(z)^2)^{3/2}} = - \frac{z}{\lambda^2} \, .
\end{align}
For contact angle $\theta=0$, the boundary conditions at the sphere, $z_1=h-R(1+\cos \beta)$, for a given angle $\beta$, Fig.\,\ref{fig_sp2}, are
\begin{align}
 \begin{split}
  r_0(z_1) &= R \sin \beta \, , \\
	r_0'(z_1) & = \cot \beta \, . \\
 \end{split}
\end{align}

One would expect, that for small spheres gravity is not significant, and one can ignore the term $-z/\lambda^2$ on the rhs of Eq.\,(\ref{eq_shape0}). Then the equilibrium shape would be $r_0(z)=w \cosh (z-z_0)/w$, with $w$, $z_0$ -- two integration constants. Note, that for angles $\beta<\pi/2$, for $z=z_0$, the bridge has minimal width equal to $w=r_0(z_0)$. The slope of the interface at $z=0$ equals $r_0'(0)=-\sinh z_0/w$, and  because $w, z_0>0$ are finite, it cannot be infinite. The condition that the position of the interface has both infinite value and slope at $z=0$ cannot be satisfied. Thus, the function $r_0(z)=w \cosh (z-z_0)/w$ is not an acceptable equilibrium shape of the bridge for our considerations. It turns out that the term $-z/\lambda^2$ in Eq.\,(\ref{eq_shape0}) provides that $r_0'(0) \to -\infty$, and it cannot be neglected.

\begin{figure}[htb] 
 \begin{center}
  \includegraphics[width= \tw]{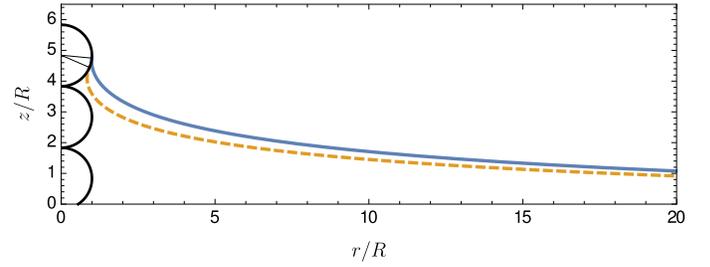}
	\caption{Shapes of equilibrium (full line) and metastable (dashed line) sphere-planar liquid surface bridges for height $h=175 \, \un{\mu m}$, radius of spheres $R=30 \, \un{\mu m}$, and capillary length $\lambda=1.45 \, \un{mm}$. \label{fig_eqmet}}
	\end{center}
\end{figure}

\begin{figure}[htb]
 \begin{center}
  \includegraphics[width= \tw]{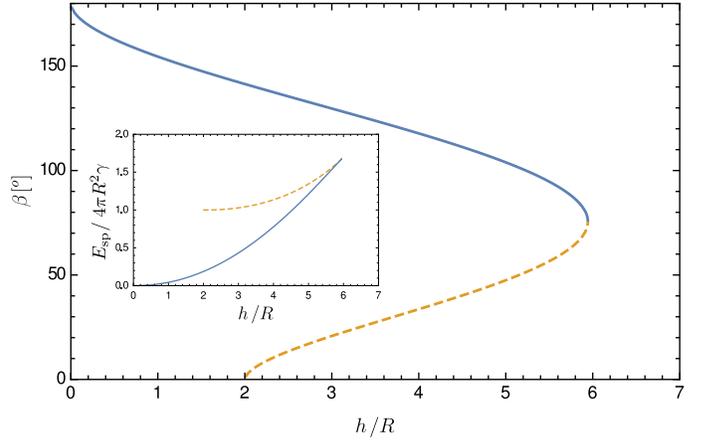}
	\caption{The plot of angle $\beta$ and the energy of a sphere-planar liquid surface bridge $E_\un{sp}$ (inset) as a function of height $h$ for equilibrium (full line) and metastable (dashed line) solutions. The sphere radius $R=30 \, \un{\mu m}$, and the capillary length $\lambda=1.45 \, \un{mm}$. \label{fig_wykb}}
	\end{center}
\end{figure}

A procedure of numerical integration of Eq.\,(\ref{eq_shape0}) is described in Appendix\,\ref{app_shape}. It turns out that there exists a particular value of the height $h=h_\un{sp}$ (spinodal height), such that for  $h<h_\un{sp}$ there exist two solutions of Eq.\,(\ref{eq_shape0}) corresponding to different angles $\beta$, Fig.\,\ref{fig_eqmet}.  On the other hand, for large heights $h> h_\un{sp}$, the solution of Eq.\,(\ref{eq_shape0}) ceases to exist. We checked that in the case $h<h_\un{sp}$ when two solutions exist, the solution corresponding to the larger value of angle $\beta$ always has lower energy than the solution corresponding to the lower value of $\beta$, Fig.\,\ref{fig_wykb}. Hence, the solution for larger angle $\beta$ describes the equilibrium capillary bridge, while the one with lower $\beta$ -- the metastable capillary bridge. For equilibrium solution, the angle $\beta$ is a decreasing function of $h$, and for metastable solution, it is an increasing function of $h$, Fig.\,\ref{fig_wykb}.

\section{Continuous and discontinuous transitions \label{sec_transitions}}
\begin{figure}[htb]
 \begin{center}
  \includegraphics[width= \tw]{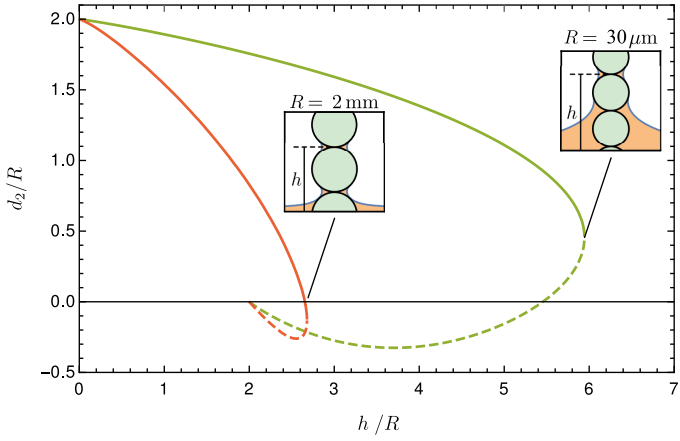}
	\caption{The distance $d_2(h)$ for equilibrium bridges (full lines) and metastable bridges (dashed lines), for spheres with radius $R=30 \, \un{\mu m}<\lambda/2$ (green), and $R=2 \, \un{mm} >\lambda/2$ (orange). The capillary length equals $\lambda=1.45 \, \un{mm}$. \label{fig_wykd2RR}}
 \end{center}
\end{figure}

We note that some of the solutions of Eq.\,(\ref{eq_shape0}) are non-physical, because when the presence of the lower, neighbouring sphere is taken into account, then the liquid-gas interface crosses this sphere. In other words, there exists $0<z<z_1$, for which $r_0(z)<r_1(z)$. In order to track the distance between the interface and the neighbouring lower sphere, the parameter 
\beq
d_2(h) = \min_{0<z<z_1} \sqrt{r_0(z)^2+(z-(h-3R))^2}-R 
\eeq
is introduced. It turns out that for $R<\lambda/2$, this distance is always positive for equilibrium bridges, while for $R>\lambda/2$, it achieves zero for certain $h_\un{tr}<h_\un{sp}$, Fig.\,\ref{fig_wykd2RR}. 

\begin{figure}[htb]
 \begin{center}
  \includegraphics[width= \tw]{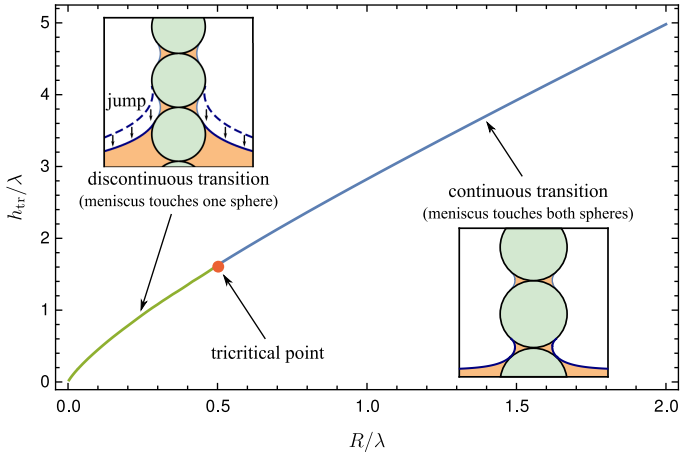}
	\caption{Plot of the height $h_\un{tr}$, at which the morphological phase transition takes place, as a function of radius of the spheres $R$ forming a chain. For $R<\lambda/2$, the transition is discontinuous, and for $R>\lambda/2$, the transition is continuous. The value $R=\lambda/2$ determines the tricritical point $(R_\un{tric}=0.5 \lambda,h_\un{tric}=1.63 \lambda)$. \label{fig_wtricrit}}
 \end{center}
\end{figure}

Thus, for large spheres ($R>\lambda/2$), a continuous morphological transition takes place at $h_\un{tr}$, such that $d_2(h_\un{tr})=0$. The sphere-planar liquid surface bridge transforms continuously into: (1) the bridge between two neighbouring spheres, and (2) the bridge between lower sphere and the planar liquid surface, ESI Animation 1. On the contrary, for small spheres ($R<\lambda/2$), the morphological transition is discontinuous. It takes place at height $h_\un{tr}=h_\un{sp}$, for which the sphere-planar liquid interface bridge ceases to exist, ESI Animation 2. At this height, the sphere-sphere bridge and sphere-planar liquid interface bridge connecting the lower sphere with the planar liquid surface form discontinuously. Thus, the particular value of the radius $R=\lambda/2$, in the $(R,h)$ space, corresponds to a tricritical point $(R_\un{tric}=0.5 \lambda,h_\un{tric}=1.63 \lambda)$, i.e. the point, where the line of discontinuous morphological phase transitions meets the line of continuous morphological phase transitions \cite{Yeomans1992}, Fig.\,\ref{fig_wtricrit}. We note, that during the process of pulling the chain out of the liquid, the above transition takes place periodically, with period $\Delta h =2R$. 

\begin{figure}[htb]
 \begin{center}
  \includegraphics[width= \tw]{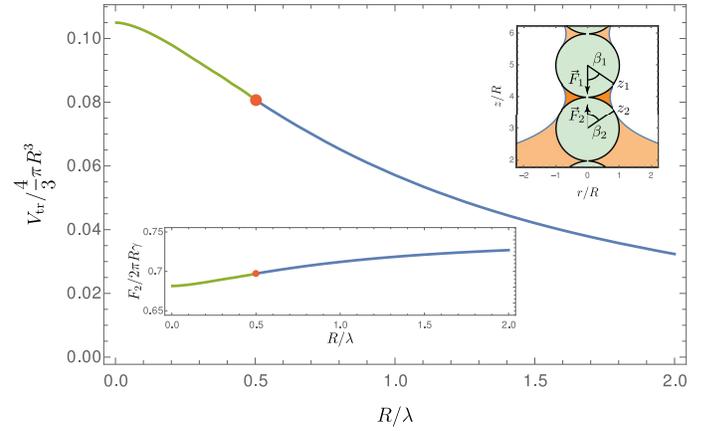}
	\caption{The volume of the sphere-sphere bridge $V_\un{tr}$ formed during the morphological phase transitions as a function of the radius of the spheres $R$. The radius $R=\lambda/2$, where $\lambda$ is a capillary length, corresponds to the tricritical point. The inset shows the capillary force $\vec{F}_2$ acting upward on the lower sphere. \label{fig_spsp_volumes}}
 \end{center}
\end{figure}

\begin{figure}[htb]
 \begin{center}
  \includegraphics[width= \tw]{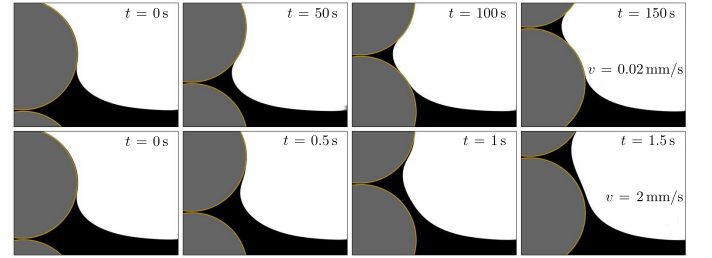}
	\caption{Two  sequences of photos illustrating the variations of the shape of the sphere-planar liquid surface bridge which take place in the process of pulling the chain of spheres out of the liquid. The upper panel corresponds to the chain vertical velocity $v=0.02 \, \un{mm/s}$, and the lower panel corresponds to $v=2 \, \un{mm/s}$. In both cases, $R=2 \, \un{mm}$. Note that  $R>\lambda/2 = 0.725 \, \un{mm}$, and we observe a continuous morphological transition in which the sphere-sphere bridge is formed. Grey circles with yellow edges were added to mark both the shape and the position of the spheres. \label{fig_Exp2}}
	\end{center}
\end{figure}

The shape of the liquid bridge between two adjacent spheres that is formed in the continuous morphological transition is described by Eq.\,(\ref{eq_shape0}). Once this bridge is formed, its volume and the shape remain unchanged, there is no liquid flow along the surface of the spheres \cite{Bayramli1987,Adams2002,Lian2016}. On the other hand, in the case of the sphere-sphere bridge formed in the discontinuous transition, we assume that its shape is such that it minimizes the surface free energy. Thus, also in this case, the shape of the bridge is described by Eq.\,(\ref{eq_shape0}). Volume of the liquid bridge between two adjacent spheres is determined at transition. It is a function of the radius of the spheres $R$, and doesn't depend on $h$, Fig.\,\ref{fig_spsp_volumes}. In situations, in which the velocity of the chain being pulled out from the liquid can not be neglected, the volume of the bridge depends on the velocity, Fig.\,\ref{fig_Exp2}.

\section{Capillary forces \label{sec_forces}}

During the process of pulling the chain of spheres out from the liquid bath, i.e., for increasing values of parameter $h$, the shape of the bridge undergoes modifications, which induces modifications of the capillary force acting downward on the chain. The capillary force $\vec{F}_\un{sp}(h) = F_\un{sp}(h) \, \vec{e}_z$, where $\vec{e}_z$ denotes the unit vector directed upwards, can be calculated (see Appendix \ref{app_force}) on the basis of the surface free energy in Eq.\,(\ref{eq_functional}) 
\begin{align} \label{eq_Fsp}
\begin{split}
 F_\un{sp}(h) = & - \frac{{\rm d}E_\un{sp}[r_0(z)]}{{\rm d} h} \\ 
        = & - 2 \pi R \gamma \sin^2 \beta - \rho_\un{l}g z_1 \,  \pi R^2 \sin^2 \beta \\ 
  &+  \rho_\un{l} g \int_0^{z_1} \dd z \, \pi r_1(z)^2 \, .
\end{split}
\end{align}
The first term describes the contribution to the total capillary force that acts at the three phase (liquid-gas-sphere) contact line. The second term is the product of hydrostatic pressure depending on the height of the bridge $z_1$ and the cross-section area of the sphere at $z=z_1$. The third term corresponds to the buoyancy force acting upwards.

\begin{figure}[htb]
 \begin{center}
  \includegraphics[width= \tw]{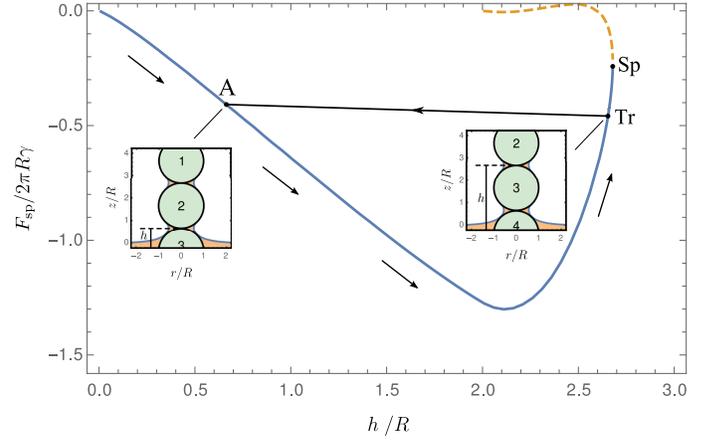}
	\caption{Capillary force in the sphere-planar liquid surface bridge $F_\un{sp}$ as a function of height $h$ for equilibrium (solid line) and metastable (dashed line) bridges. Black solid line joins the morphological transition point $\un{Tr}$, of height $h=h_\un{tr}$, with the point $\un{A}$, of height $h_\un{tr}-2R$, where the new period begins. The spinodal point $\un{Sp}$ denotes the height, above which, in the system without the presence of the second lower sphere, the bridge cease to exist. Radius of the sphere is $R=2 \, \un{mm} > \lambda/2$, where capillary length equals $\lambda=1.45 \, \un{mm}$. In the case of continuous transition, the difference $F_\un{sp}(h_\un{tr}-2R)-F_\un{sp}(h_\un{tr})$ equals the weight of the liquid bridge. \label{fig_FspR2}}
 \end{center}
\end{figure}

In the case of the sphere-sphere bridge, one can distinguish two forces \cite{Adams2002}, $\vec{F}_1 = F_1 \, \vec{e}_z$ and $\vec{F}_2 = F_2 \, \vec{e}_z$. Force $\vec{F}_1$ acts on the upper sphere, is directed downward, and
\begin{align}
\begin{split}
F_1 = & \, - 2 \pi R \gamma \sin^2 \beta_1 - \rho_\un{l}g \, z_1 \pi R^2 \sin^2 \beta_1 \\
 & +\rho_\un{l} g \int_{h-2R}^{z_1} \dd \, z \pi r_1(z)^2 \, .
\end{split}
\end{align}
Force $\vec{F}_2$ acts on the lower sphere, is directed upward, and
\begin{align}
\begin{split}
F_2 =& \, 2 \pi R \gamma \sin^2 \beta_2 + \rho_\un{l}g \, z_2 \pi R^2 \sin^2 \beta_2 \\
 & +\rho_\un{l} g \int_{z_2}^{h-2R} \dd z \, \pi r_1(z)^2 \, ,
\end{split}
\end{align}
where $\beta_{2}$ is the angle between the vertical direction and the radius directed to the three phase contact line, which is located at  height $z_{2}$ (see inset in Fig.\,\ref{fig_spsp_volumes}). Both these forces have the same structure as the capillary force acting between the sphere and the planar liquid surface, Eq.\,(\ref{eq_Fsp}). For the continuous morphological transition ($R>\lambda/2$), one can check that
\beq
 F_\un{sp}(h_\un{tr}) = F_\un{sp}(h_\un{tr}-2R)+F_1+F_2 \, .
\eeq
On the other hand, the sum of the forces for the sphere-sphere bridge case equals the weight of the liquid bridge (see Appendix \ref{app_volume})
\beq
 F_1+F_2 = -\rho_l g V_\un{tr} \, ,
\eeq 
where $V_\un{tr}$ denotes the volume of the sphere-sphere bridge. Finally, we obtain 
\beq \label{Ftr}
 F_\un{sp}(h_\un{tr}) = F_\un{sp}(h_\un{tr}-2R)-\rho_\un{l} g V_\un{tr} \, .
\eeq
Thus, in the case $R>\lambda/2$ (and for very small velocities), one can obtain the volume of the sphere-sphere bridge from the capillary force measurements, Fig.\,\ref{fig_FspR2}.    

\begin{figure}[htb]
 \begin{center}
  \includegraphics[width= \tw]{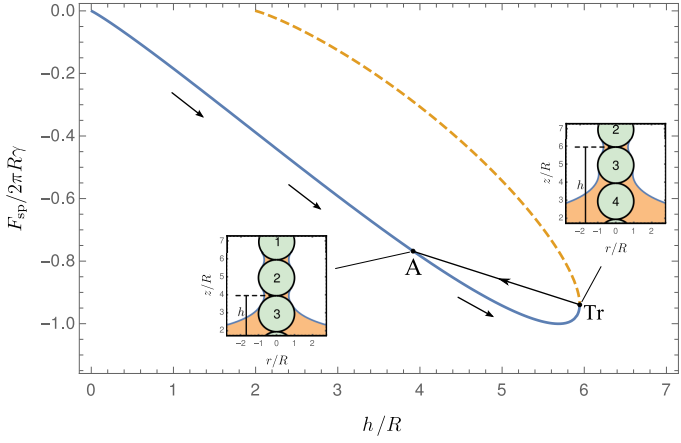}
	\caption{Capillary force in the sphere-planar liquid surface bridge $F_\un{sp}$ as a function of height $h$ for equilibrium (solid line) and metastable (dashed line) bridges. Black solid line joins the morphological transition point $\un{Tr}$, of height $h=h_\un{tr}$, with the point $\un{A}$, of height $h_\un{tr}-2R$, where the new period begins. The weight of the sphere-sphere bridge $\rho_\un{l} g V_\un{tr}/2 \pi R \gamma = 3 \cdot 10^{-4}$ doesn't compensate the difference  $F_\un{sp}(h_\un{tr}-2R)-F_\un{sp}(h_\un{tr})$, and one observes discontinuity in the force acting on the chain. Radius of the sphere is $R=30 \, \un{\mu m} < \lambda/2$, where capillary length equals $\lambda=1.45 \, \un{mm}$. \label{fig_FspR003}}
 \end{center}
\end{figure}

In the case of discontinuous transitions (small radii), the weight of the sphere-sphere bridge doesn't compensate the difference between the sphere-planar liquid interface forces that pop up during the transition 
\beq
 F_\un{sp}(h_\un{tr}) < F_\un{sp}(h_\un{tr}-2R)-\rho_\un{l} g V_\un{tr} \, , 
\eeq
and one observes discontinuity in the force acting on the chain. For example, for $R=30 \, \un{\mu m}$ and $\lambda=1.45 \, \un{mm}$, the weight of the sphere-sphere bridge $\rho_\un{l} g V_\un{tr}/2 \pi R \gamma = 3 \cdot 10^{-4}$, and it is three orders of magnitude smaller than the difference between the sphere-planar liquid interface forces, Fig.\,\ref{fig_FspR003}. 

We note, that the absolute value of the capillary force acting upward (see inset in Fig.\,\ref{fig_spsp_volumes}), is smaller than the maximum of the absolute value 
of sphere-liquid surface bridge $|F_2|<\max_{h} |F_\un{sp}|$. Thus, in the case a chain is formed by assembling spheres from a suspension (as for example in the assembly route described in reference \cite{Rozynek2016}), one needs, on top of the capillary force, an additional force of attraction acting between the spheres (e.g., the dipolar force).

\section{Comparison with experiment \label{sec_experiment}}

\begin{figure}[htb]
 \begin{center}
  \includegraphics[width= \tw]{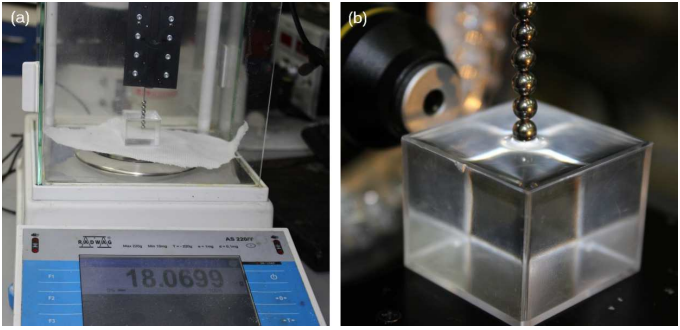}
	\caption{The experimental setup: (a) A chain of glued steel spheres of radii $R=2 \, \un{mm}$ is being pulled out from $10 \, \un{cSt}$ silicone oil by a stepper motor with constant velocity. Weight difference is measured by microscale during experiment; (b) Close-up on the spheres, silicone oil bath and the camera recording the experiment. \label{fig_Exp1}}
	\end{center}
\end{figure}

To check the validity of our model, we prepared a chain of spheres of radius $R=2 \, \un{mm}$ ($R>\lambda/2$, where $\lambda=1.45 \, \un{mm}$). The adjacent spheres in the chain were glued together. The chain was very slowly  pulled out from a $10 \, \un{cSt}$ silicone oil bath with $v=0.02 \, \un{mm/s}$ to ensure quasistatic conditions. A microscale was used as a very precise dynamometer, Fig.\,\ref{fig_Exp1}. 

\begin{figure}[htb]
 \begin{center}
  \includegraphics[width= \tw]{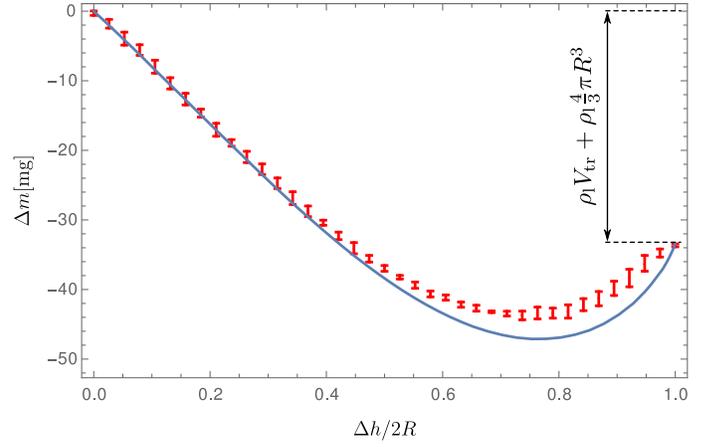}
	\caption{Comparison of theoretical predictions and experimental measurements of the mass difference $\Delta m$ during the process of pulling out $R=2 \, \un{mm}$ steel spheres on height $\Delta h = h-(h_\un{tr}-2R)$ from a $10 \, \un{cSt}$ silicone oil bath. For $\Delta h = 2R$ the absolute value of the mass difference equals sum of mass of one sphere-sphere liquid bridge, and the mass of liquid displaced by one sphere, $\Delta m (2R) = - 33.27 \, \un{mg}$. Measurements were averaged over four spheres, and for theoretical model calculations the capillary tension was taken to be $\gamma=19.7 \, \un{mN/m}$, oil density $\rho_\un{l}=950 \, \un{kg/m^3}$, and the maximum radius of the system in numerical calculations to $r_\un{max}=10 \, \un{mm}$. \label{fig_thexp}}
	\end{center}
\end{figure}

For a given height $h$, weight on the microscale $m(h) \, g$ was equal to the weight of the silicone oil reduced by: the weight of the liquid in the sphere-sphere bridges, the absolute value of the sphere-planar liquid surface capillary force $|F_\un{sp}(h)|$, and the buoyancy force. The buoyancy force, during the process of pulling out the spheres, decreases by a factor proportional to the volume of the spheres drawn above $z=0$ level
\beq
 \Delta F_\un{b} (h) &=& - \rho_\un{l} \, g \int_0^{h} \dd z \, \pi r_1(z)^2 \, .
\eeq  
Thus, the mass difference $\Delta m(\Delta h) = m(h)- m(h_\un{tr}-2R)$, where $\Delta h = h-(h_\un{tr}-2R)$, equals
\begin{align}
 \begin{split}
 \Delta m(\Delta h) 
   = \frac{1}{g} \Big(F_\un{sp}(h)-F_\un{sp}(h_\un{tr}-2R)\Big) - \rho_\un{l} \int_{h_\un{tr}-2R}^{h} \dd z \, \pi r_1(z)^2 \, .
\end{split}
\end{align}
After one period, using Eq.\,(\ref{Ftr}), we get 
\begin{align}
 \Delta m(2R) = -\rho_\un{l} V_\un{tr} -  \rho_\un{l} \frac{4}{3} \pi R^3 \, .
\end{align}
The first term is the mass of one sphere-sphere liquid bridge, and the second term is the mass of liquid displaced by one sphere. The comparison of experimental data and the theoretical curve is shown in Fig.\,\ref{fig_thexp}. We notice that the agreement is particularly satisfying for $\Delta h/2R < 0.5$. 


\section{Summary \label{sec_summary}}

We described theoretically the mechanism of capillary bridge formation on a beaded chain pulled out from a liquid. Two types of capillary bridges come into play. The first type is the bridge connecting the sphere with a planar liquid surface, and in the process of pulling out the chain of spheres from the liquid, this bridge appears first. Then, when the next sphere is pulled out from the liquid, this bridge transforms into the bridge between the adjacent spheres and the bridge connecting the lower sphere with the surface of the liquid. We showed that this morphological transition changes its order, depending on the ratio of the sphere radius and the capillary length $R/\lambda$. For $R/\lambda > 2$, it is continuous, and for $R/\lambda <2$, it is discontinuous with the particular value $R/\lambda =2$ corresponding to the tricritical point. The shape of the meniscus of the bridge is given by the solution of Eq.\,(\ref{eq_shape0}), in which the mean curvature on its lhs depends on the local height of the meniscus. There are two solutions of this equation, and the one corresponding to the larger value of $\beta$ always has smaller surface free energy. The metastable bridge corresponding to the smaller angle $\beta$ could be observed in the reverse experiment, when a sphere is pushed towards a flat liquid surface. 

Besides the shape of the bridges, the accompanying capillary forces acting in the system were also calculated. It turns out that the constant capillary force binding two adjacent spheres is not sufficiently strong to prevent the breaking of the chain in the process of pulling it out from the liquid. It is smaller than  the maximal value of the sphere-planar liquid surface capillary force acting downwards. This maximal value is attained for the height smaller than the height of morphological phase transition. Thus, an additional attractive force acting between the adjacent spheres is needed to provide the stability of the chain. This can be the dipolar force \cite{Rozynek2016}; in the reported experiment, the spheres were simply glued together. 

We compared our theoretical predictions with experimental data corresponding to the case $R>\lambda/2$. The observed morphological transition is continuous, as expected on theoretical grounds, and the plot of the theoretically predicted capillary force fits well the experimental data, in particular for larger $R/\lambda$ values. 

In our theoretical model, we assumed zero contact angle. Relaxing this constraint can lead to stabilization of the metastable bridges in the $\theta=0$ case. Such a change of stability would cause the phase diagram and the formation of liquid bridges scenarios to be more complicated. This is left for further analysis. 

Summarizing, we have explained theoretically the mechanisms of capillary bridges formation in the process of  pulling the chain of spheres out of a liquid bath. We predicted the existence of a morphological phase transition between two types of bridges, which can be continuous or discontinuous. The transition is continuous when the diameter of the spheres is larger than the capillary length. In the opposite case the transition is discontinuous, and so is the capillary force acting on the chain. Thus the capillary length sets the lower limit for the diameter of the spheres, for which the beaded chain formation out from a liquid dispersion is a smooth process. 

\appendix
\section{Procedure of finding the shape of liquid bridges \label{app_shape}}

In order to find the shapes of liquid bridges, we solve Eq.\,(\ref{eq_shape0}) numerically using the shooting method \cite{Dutka2007,Numerical2007}.  The essential ingredient of this method consists of checking whether the boundary condition $r_0(z=0)=\infty$ is fulfilled for the probe function at hand.  In our numerical calculations, we check this condition using a large but finite parameter denoted as $r_\un{N}$.  We also introduce another auxiliary quantity $z_\un{N}$. It is defined as the minimal value of $z$ in the range $[0, h-R(1+\cos \beta)]$, for which $r_{0}(z_\un{N}) \leqslant r_\un{N}$. Such defined quantity $z_\un{N}$ depends on the angle $\beta$, i.e., we have $z_\un{N}(\beta)$. Note that if $r_{0}(0) < r_\un{N}$, then we adjust the angle $\beta$ to increase the value of $r_{0}$. 

\begin{figure}[htb]
 \begin{center}
  \includegraphics[width = \tw]{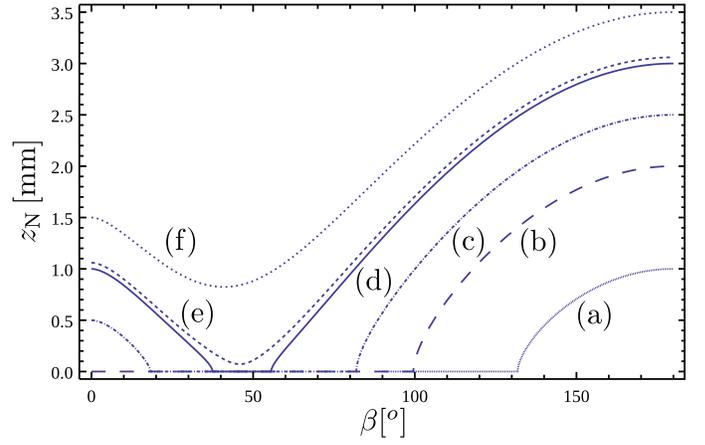}
	\caption{Function $z_\un{N}(\beta)$ for different heights: (a) $h = 1 \, \un{mm}$, (b) $h = 2 \, \un{mm}$, (c) $h = 2.5 \, \un{mm}$, (d) $h = 3 \, \un{mm}$, (e) $h = 3.06 \, \un{mm}$, and (f) $h = 3.5 \, \un{mm}$. Radius of a sphere is $R=1 \, \un{mm}$, and capillary length $\lambda = 1.45 \, \un{mm}$. For $h \leqslant 2R$, function $z_\un{N}(\beta)=0$ between zero and certain value of $\beta$, and then it is increasing up to $\beta=180^\un{o}$. In these calculations, we took $r_\un{N} = 10^6 R$. \label{fig_zN}}
 \end{center}
\end{figure}

The equilibrium profile corresponds to $z_\un{N}=0$. Function $z_\un{N}(\beta)$ is non-differentiable (has a kink), Fig.\,\ref{fig_zN}. Depending on the value of $h$, the equation $z_\un{N}(\beta)=0$ can have zero, one, or two solutions, at which points the function $z_\un{N}(\beta)$ is non-differentiable. Every such point corresponds to solution of Eq.\,(\ref{eq_shape0}). For $h>h_\un{tr}$, we have no solution, for $h = h_\un{tr}$ -- one solution, for $2R < h < h_\un{tr}$ -- two solutions, and for $0<h<2R$ -- one solution. While two solutions of Eq.\,(\ref{eq_shape0}) are possible, the equilibrium shape of the bridge corresponds to the one with smaller surface energy. We checked, that the solution with  larger $\beta$ always has smaller surface energy. 

\begin{figure}[htb]
 \begin{center}
  \includegraphics[width = \tw]{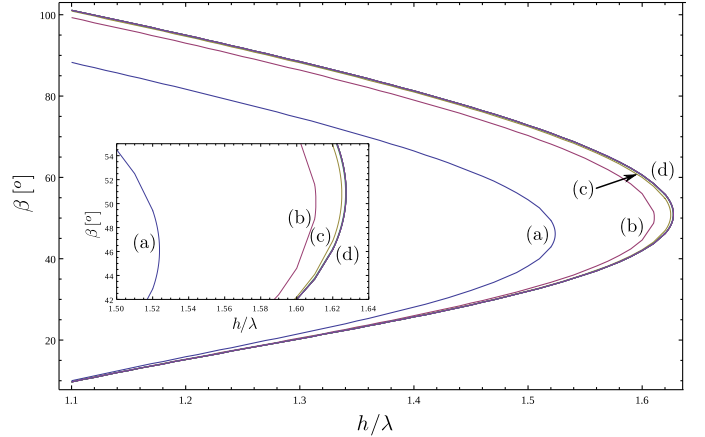}
	\caption{Dependence of $\beta$ on $h$ for equilibrium (upper branch) and metastable (lower branch) solutions for different choices of $r_\un{N}$: (a) $r_\un{N} = \lambda$, (b) $r_\un{N} = 2 \lambda$, (c) $r_\un{N} = 3\lambda$, (d) $r_\un{N} = 4 \lambda, 5 \lambda, 6 \lambda, 7 \lambda, 8 \lambda, 9 \lambda,  10 \lambda$. The inset shows close-up of transition regions. In all the calculations, $R=\lambda/2$.
	\label{fig_rN}}
 \end{center}
\end{figure}

To check how the choice of $r_\un{N}$ influences the results, we performed calculations of $\beta(h)$ for different $r_\un{N}$ values, Fig.\,\ref{fig_rN}. We note that starting from $r_\un{N}>4 \lambda$, the results do not change significantly, so in the following calculations we will  take $r_\un{N}$ to be in the vicinity of $10 \lambda$. 

\section{Calculation of the capillary force \label{app_force}}

The functional in Eq.\,(\ref{eq_functional}) can be rewritten, using Heaviside step function $\Theta(z)$ and Dirac delta function $\delta(z)$, in the following form 
\begin{align} \label{eq_functional0}
 \begin{split}
\frac{E_\un{sp}[r(z)]}{2 \pi \gamma} = & \int \dd z  \Bigg[\Bigg( r(z) \sqrt{1+r'(z)^2} 
	 	  +  \frac{z}{2 \lambda^2} \Big(r(z)^2-r_1(z)^2 \Big)-R \Bigg) \\ &\qquad \qquad  \Theta\Big(r(z)-r_1(z)\Big) + R \Bigg] \Theta(z) \Theta(h-z) \\
	& -  \int \dd z \, \delta(z) r(z)^2/2  \, .
\end{split}
\end{align}

The sphere-liquid planar surface capillary force $F_\un{sp}$ is then given by 
\begin{align}
\begin{split}
 - \frac{F_\un{sp}}{2 \pi \gamma} &=  \frac{1}{2 \pi \gamma} \frac{{\rm d}E_\un{sp}[r_0(z)]}{{\rm d} h} \\ 
 &= \frac{1}{2 \pi \gamma} \Bigg( \left.\frac{\delta E_\un{sp}[r(z)]}{\delta r(z)}\right|_{r=r_0(z)} \frac{{\rm d} r_0(z)}{{\rm d} h}
  + \frac{\partial E_\un{sp}[r_0(z)]}{\partial h} \Bigg)  \\
	&=  R + \int \dd z \Bigg[ \Bigg( r_0(z) \sqrt{1+r_0'(z)^2} - R \Bigg)  \frac{\delta(z-z_1)}{|r_0'(z_1)-r_1'(z_1)|} \\
	& 	\qquad \qquad \qquad  +  \frac{z}{\lambda^2} r_1(z) \Bigg] \Big(- \frac{{\rm d} r_1(z)}{{\rm d} h} \Big)  \\
	&=  R + \int \dd z \Bigg[ \Bigg( r_0(z) \sqrt{1+r_0'(z)^2} - R \Bigg)  \frac{\delta(z-z_1)}{|r_0'(z_1)-r_1'(z_1)|} \\
	 & 	\qquad \qquad \qquad  	  +  \frac{z}{\lambda^2} r_1(z) \Theta \Big(r_0(z)-r_1(z)\Big)\Theta(z) \Bigg)\Bigg]  \frac{{\rm d}r_1(z)}{{\rm d} z}   \\
	&=	R + \cot \beta \frac{r_0(z_1) \sqrt{1+r_0'(z_1)^2} - R}{|r_0'(z_1)-r_1'(z_1)|} + \frac{1}{2 \lambda^2}\int_0^{z_1} \dd z \, z \frac{{\rm d} (r_1(z)^2)}{{\rm d} z}  \\
	&= R \sin^2 \beta + \frac{z_1}{\lambda^2} \frac{r_1(z_1)^2}{2} -  \frac{1}{2 \lambda^2} \int_0^{z_1} \dd z \, r_1(z)^2 \, ,
\end{split}
\end{align}
where $r_0(z)$ is the equilibrium shape of the sphere-planar liquid surface bridge interface. Because $r_1(z)$ depends on $h$ only through the difference $z-h$, then $r_1(z) = r_\un{s}(z-h)$, ${\rm d} r_1(z) / {\rm d} h = -{\rm d} r_1(z) / {\rm d} z$, where $r_\un{s}(z)$ is a function describing submerged spheres ($h=0$). To calculate $(r_0(z_1) \sqrt{1+r_0'(z_1)^2} - R)/|r_0'(z_1)-r_1'(z_1)|$, we assumed that $r_0'(z_1) = \cot (\beta+\theta)$, and then took the limit $\theta \to 0$.

\section{Volume of the bridge \label{app_volume}}
To find the volume of the sphere-sphere bridge, one can integrate once Eq.\,(\ref{eq_shape0}) to get
\begin{align}
\begin{split}
\frac{{\rm d}}{{\rm d}z} \frac{r_0(z)}{(1+r_0'(z)^2)^{1/2}} = & \, \frac{r_0'(z)}{(1+r'_0(z)^2)^{1/2}} - \frac{r_0(z) r_0'(z) r''_0(z)}{(1+r'_0(z)^2)^{3/2}} \\
  = &\,  -\frac{z}{2 \lambda^2}  \frac{{\rm d}}{{\rm d}z} (r_0(z))^2 \, ,
\end{split}
\end{align}
and hence
\begin{align}
\begin{split}
\left. \frac{r_0(z)}{\sqrt{1+r_0'(z)^2}} \right|_{z_2}^{z_1} = -\left. \frac{z r_0(z)^2}{2 \lambda^2}\right|_{z_2}^{z_1} + \frac{1}{2 \lambda^2} \int_{z_2}^{z_1} \dd z r_0(z)^2 \, ,
\end{split}
\end{align}
which gives
\begin{align}
\begin{split}
V_\un{tr} = & \int_{z_2}^{z_1} \dd z \, \pi (r_0(z)^2 - r_1(z)^2) \\
  =& \,  2 \pi \lambda^2 R \Big(\sin^2 \beta_1 - \sin^2 \beta_2 \Big) \\
   & \, + \pi R^2 \Big(z_1  \sin^2 \beta_1 - z_2 \sin^2 \beta_2 \Big) 
 - \int_{z_2}^{z_1} \dd z \, \pi  r_1(z)^2 \\
 = & - \frac{1}{\rho_\un{l} g} \Big(F_1+F_2 \Big)\, .
\end{split}
\end{align}
The total capillary force acting on two adjacent spheres is equal to the weight of the liquid bridge between them \cite{Adams2002}.

\section*{Acknowledgments}
F.D. was supported by the Foundation for Polish Science, Poland, within the project Homing Plus/2012-6/3, co-financed from the European Regional Development Fund. Z.R. acknowledges financial support from the National Science Centre, Poland, through the OPUS programme (2015/19/B/ST3/03055) and the Foundation for Polish Science, through the Homing Plus programme (2013-7/13). The authors are grateful to Dr Jan Guzowski for numerous comments and suggestions.


 \balance


\bibliography{bibliography}

\providecommand*{\mcitethebibliography}{\thebibliography}
\csname @ifundefined\endcsname{endmcitethebibliography}
{\let\endmcitethebibliography\endthebibliography}{}
\begin{mcitethebibliography}{71}
\providecommand*{\natexlab}[1]{#1}
\providecommand*{\mciteSetBstSublistMode}[1]{}
\providecommand*{\mciteSetBstMaxWidthForm}[2]{}
\providecommand*{\mciteBstWouldAddEndPuncttrue}
  {\def\EndOfBibitem{\unskip.}}
\providecommand*{\mciteBstWouldAddEndPunctfalse}
  {\let\EndOfBibitem\relax}
\providecommand*{\mciteSetBstMidEndSepPunct}[3]{}
\providecommand*{\mciteSetBstSublistLabelBeginEnd}[3]{}
\providecommand*{\EndOfBibitem}{}
\mciteSetBstSublistMode{f}
\mciteSetBstMaxWidthForm{subitem}
{(\emph{\alph{mcitesubitemcount}})}
\mciteSetBstSublistLabelBeginEnd{\mcitemaxwidthsubitemform\space}
{\relax}{\relax}

\bibitem[Balakin \emph{et~al.}(2015)Balakin, Kutsenko, Layrukhin, and
  Kosinski]{Balakin2015}
B.~V. Balakin, K.~V. Kutsenko, A.~A. Layrukhin and P.~Kosinski, The collision
  efficiency of liquid bridge agglomeration, \emph{Chemical Engineering
  Science}, 2015, \textbf{137}, 590--600\relax
\mciteBstWouldAddEndPuncttrue
\mciteSetBstMidEndSepPunct{\mcitedefaultmidpunct}
{\mcitedefaultendpunct}{\mcitedefaultseppunct}\relax
\EndOfBibitem
\bibitem[Wang \emph{et~al.}(2015)Wang, Li, Liu, and Zhao]{Wang2015}
W.~Wang, Y.~Li, H.~Liu and P.~Zhao, Study of {Agglomeration} {Characteristics}
  of {Hydrate} {Particles} in {Oil}/{Gas} {Pipelines}, \emph{Advances in
  Mechanical Engineering}, 2015, \textbf{7}, 457050\relax
\mciteBstWouldAddEndPuncttrue
\mciteSetBstMidEndSepPunct{\mcitedefaultmidpunct}
{\mcitedefaultendpunct}{\mcitedefaultseppunct}\relax
\EndOfBibitem
\bibitem[Balakin \emph{et~al.}(2013)Balakin, Alyaev, Hoffmann, and
  Kosinski]{Balakin2013}
B.~V. Balakin, S.~Alyaev, A.~C. Hoffmann and P.~Kosinski, Micromechanics of
  agglomeration forced by the capillary bridge: {The} restitution of momentum,
  \emph{AIChE Journal}, 2013, \textbf{59}, 4045--4057\relax
\mciteBstWouldAddEndPuncttrue
\mciteSetBstMidEndSepPunct{\mcitedefaultmidpunct}
{\mcitedefaultendpunct}{\mcitedefaultseppunct}\relax
\EndOfBibitem
\bibitem[Herminghaus(2005)]{Herminghaus2005}
S.~Herminghaus, Dynamics of wet granular matter, \emph{Advances in Physics},
  2005, \textbf{54}, 221--261\relax
\mciteBstWouldAddEndPuncttrue
\mciteSetBstMidEndSepPunct{\mcitedefaultmidpunct}
{\mcitedefaultendpunct}{\mcitedefaultseppunct}\relax
\EndOfBibitem
\bibitem[Scheel \emph{et~al.}(2008)Scheel, Seemann, Brinkmann, Di~Michiel,
  Sheppard, Breidenbach, and Herminghaus]{Sheel2008}
M.~Scheel, R.~Seemann, M.~Brinkmann, M.~Di~Michiel, A.~Sheppard, B.~Breidenbach
  and S.~Herminghaus, Morphological clues to wet granular pile stability,
  \emph{Nature Materials}, 2008, \textbf{7}, 189--193\relax
\mciteBstWouldAddEndPuncttrue
\mciteSetBstMidEndSepPunct{\mcitedefaultmidpunct}
{\mcitedefaultendpunct}{\mcitedefaultseppunct}\relax
\EndOfBibitem
\bibitem[Pakpour \emph{et~al.}(2012)Pakpour, Habibi, M{\o}ller, and
  Bonn]{Pakpour2012}
M.~Pakpour, M.~Habibi, P.~M{\o}ller and D.~Bonn, How to construct the perfect
  sandcastle, \emph{Scientific Reports}, 2012, \textbf{2}, 549\relax
\mciteBstWouldAddEndPuncttrue
\mciteSetBstMidEndSepPunct{\mcitedefaultmidpunct}
{\mcitedefaultendpunct}{\mcitedefaultseppunct}\relax
\EndOfBibitem
\bibitem[Xue \emph{et~al.}(2015)Xue, Kovalev, Eichler-Volf, Steinhart, and
  Gorb]{Xue2015}
L.~Xue, A.~Kovalev, A.~Eichler-Volf, M.~Steinhart and S.~N. Gorb,
  Humidity-enhanced wet adhesion on insect-inspired fibrillar adhesive pads,
  \emph{Nature Communications}, 2015, \textbf{6}, 6621\relax
\mciteBstWouldAddEndPuncttrue
\mciteSetBstMidEndSepPunct{\mcitedefaultmidpunct}
{\mcitedefaultendpunct}{\mcitedefaultseppunct}\relax
\EndOfBibitem
\bibitem[Slater \emph{et~al.}(2014)Slater, Vogel, Macner, and
  Steen]{Slater2014}
D.~M. Slater, M.~J. Vogel, A.~M. Macner and P.~H. Steen, Beetle-inspired
  adhesion by capillary-bridge arrays: pull-off detachment, \emph{Journal of
  Adhesion Science and Technology}, 2014, \textbf{28}, 273--289\relax
\mciteBstWouldAddEndPuncttrue
\mciteSetBstMidEndSepPunct{\mcitedefaultmidpunct}
{\mcitedefaultendpunct}{\mcitedefaultseppunct}\relax
\EndOfBibitem
\bibitem[Koos and Willenbacher(2011)]{Koos2011}
E.~Koos and N.~Willenbacher, Capillary {Forces} in {Suspension} {Rheology},
  \emph{Science}, 2011, \textbf{331}, 897--900\relax
\mciteBstWouldAddEndPuncttrue
\mciteSetBstMidEndSepPunct{\mcitedefaultmidpunct}
{\mcitedefaultendpunct}{\mcitedefaultseppunct}\relax
\EndOfBibitem
\bibitem[Koos and Willenbacher(2012)]{Koos2012}
E.~Koos and N.~Willenbacher, Particle configurations and gelation in capillary
  suspensions, \emph{Soft Matter}, 2012, \textbf{8}, 3988--3994\relax
\mciteBstWouldAddEndPuncttrue
\mciteSetBstMidEndSepPunct{\mcitedefaultmidpunct}
{\mcitedefaultendpunct}{\mcitedefaultseppunct}\relax
\EndOfBibitem
\bibitem[Hoffmann \emph{et~al.}(2014)Hoffmann, Koos, and
  Willenbacher]{Hoffmann2014}
S.~Hoffmann, E.~Koos and N.~Willenbacher, Using capillary bridges to tune
  stability and flow behavior of food suspensions, \emph{Food Hydrocolloids},
  2014, \textbf{40}, 44--52\relax
\mciteBstWouldAddEndPuncttrue
\mciteSetBstMidEndSepPunct{\mcitedefaultmidpunct}
{\mcitedefaultendpunct}{\mcitedefaultseppunct}\relax
\EndOfBibitem
\bibitem[Fan \emph{et~al.}(2015)Fan, Wang, Rong, and Sun]{Fan2015}
Z.~Fan, L.~Wang, W.~Rong and L.~Sun, Dropwise condensation on a hydrophobic
  probe-tip for manipulating micro-objects, \emph{Applied Physics Letters},
  2015, \textbf{106}, 084105\relax
\mciteBstWouldAddEndPuncttrue
\mciteSetBstMidEndSepPunct{\mcitedefaultmidpunct}
{\mcitedefaultendpunct}{\mcitedefaultseppunct}\relax
\EndOfBibitem
\bibitem[Vasudev and Zhe(2008)]{Vasudev2008}
A.~Vasudev and J.~Zhe, A capillary microgripper based on electrowetting,
  \emph{Applied Physics Letters}, 2008, \textbf{93}, 103503\relax
\mciteBstWouldAddEndPuncttrue
\mciteSetBstMidEndSepPunct{\mcitedefaultmidpunct}
{\mcitedefaultendpunct}{\mcitedefaultseppunct}\relax
\EndOfBibitem
\bibitem[Arutinov \emph{et~al.}(2014)Arutinov, Smits, Albert, Lambert, and
  Mastrangeli]{Arutinov2014}
G.~Arutinov, E.~C.~P. Smits, P.~Albert, P.~Lambert and M.~Mastrangeli,
  In-{Plane} {Mode} {Dynamics} of {Capillary} {Self}-{Alignment},
  \emph{Langmuir}, 2014, \textbf{30}, 13092--13102\relax
\mciteBstWouldAddEndPuncttrue
\mciteSetBstMidEndSepPunct{\mcitedefaultmidpunct}
{\mcitedefaultendpunct}{\mcitedefaultseppunct}\relax
\EndOfBibitem
\bibitem[Broesch \emph{et~al.}(2014)Broesch, Shiang, and
  Frechette]{Broesch2014}
D.~J. Broesch, E.~Shiang and J.~Frechette, Role of substrate aspect ratio on
  the robustness of capillary alignment, \emph{Applied Physics Letters}, 2014,
  \textbf{104}, 081605\relax
\mciteBstWouldAddEndPuncttrue
\mciteSetBstMidEndSepPunct{\mcitedefaultmidpunct}
{\mcitedefaultendpunct}{\mcitedefaultseppunct}\relax
\EndOfBibitem
\bibitem[Mastrangeli(2015)]{Mastrangeli2015}
M.~Mastrangeli, The {Fluid} {Joint}: {The} {Soft} {Spot} of {Micro}- and
  {Nanosystems}, \emph{Advanced Materials}, 2015, \textbf{27}, 4254--4272\relax
\mciteBstWouldAddEndPuncttrue
\mciteSetBstMidEndSepPunct{\mcitedefaultmidpunct}
{\mcitedefaultendpunct}{\mcitedefaultseppunct}\relax
\EndOfBibitem
\bibitem[Velankar(2015)]{Velankar2015}
S.~S. Velankar, A non-equilibrium state diagram for liquid/fluid/particle
  mixtures, \emph{Soft Matter}, 2015, \textbf{11}, 8393--8403\relax
\mciteBstWouldAddEndPuncttrue
\mciteSetBstMidEndSepPunct{\mcitedefaultmidpunct}
{\mcitedefaultendpunct}{\mcitedefaultseppunct}\relax
\EndOfBibitem
\bibitem[Dittmann \emph{et~al.}(2015)Dittmann, Maurath, Bitsch, and
  Willenbacher]{Dittmann2015}
J.~Dittmann, J.~Maurath, B.~Bitsch and N.~Willenbacher, Highly {Porous}
  {Materials} with {Unique} {Mechanical} {Properties} from {Smart} {Capillary}
  {Suspensions}, \emph{Advanced Materials}, 2015, \textbf{28}, 1689--1696\relax
\mciteBstWouldAddEndPuncttrue
\mciteSetBstMidEndSepPunct{\mcitedefaultmidpunct}
{\mcitedefaultendpunct}{\mcitedefaultseppunct}\relax
\EndOfBibitem
\bibitem[Schneider \emph{et~al.}(2016)Schneider, Koos, and
  Willenbacher]{Schneider2016}
M.~Schneider, E.~Koos and N.~Willenbacher, Highly conductive, printable pastes
  from capillary suspensions, \emph{Scientific Reports}, 2016, \textbf{6},
  31367\relax
\mciteBstWouldAddEndPuncttrue
\mciteSetBstMidEndSepPunct{\mcitedefaultmidpunct}
{\mcitedefaultendpunct}{\mcitedefaultseppunct}\relax
\EndOfBibitem
\bibitem[Park \emph{et~al.}(2012)Park, Dang, Sung, and Seo]{Park2012}
K.~S. Park, J.~M. Dang, M.~M. Sung and S.-m. Seo, One-step fabrication of
  nanowire-grid polarizers using liquid-bridge-mediated nanotransfer molding,
  \emph{Nanoscale Research Letters}, 2012, \textbf{7}, 1\relax
\mciteBstWouldAddEndPuncttrue
\mciteSetBstMidEndSepPunct{\mcitedefaultmidpunct}
{\mcitedefaultendpunct}{\mcitedefaultseppunct}\relax
\EndOfBibitem
\bibitem[Wang and McCarthy(2015)]{Wang2015b}
L.~Wang and T.~J. McCarthy, Capillary-bridge-derived particles with negative
  {Gaussian} curvature, \emph{Proceedings of the National Academy of Sciences},
  2015, \textbf{112}, 2664--2669\relax
\mciteBstWouldAddEndPuncttrue
\mciteSetBstMidEndSepPunct{\mcitedefaultmidpunct}
{\mcitedefaultendpunct}{\mcitedefaultseppunct}\relax
\EndOfBibitem
\bibitem[Tisserant \emph{et~al.}(2015)Tisserant, Reissner, Beyer, Fedoryshyn,
  and Stemmer]{Tisserant2015}
J.-N. Tisserant, P.~A. Reissner, H.~Beyer, Y.~Fedoryshyn and A.~Stemmer,
  Water-{Mediated} {Assembly} of {Gold} {Nanoparticles} into {Aligned}
  {One}-{Dimensional} {Superstructures}, \emph{Langmuir}, 2015, \textbf{31},
  7220--7227\relax
\mciteBstWouldAddEndPuncttrue
\mciteSetBstMidEndSepPunct{\mcitedefaultmidpunct}
{\mcitedefaultendpunct}{\mcitedefaultseppunct}\relax
\EndOfBibitem
\bibitem[Chaix \emph{et~al.}(2006)Chaix, Gourgon, Landis, Perret, Fink,
  Reuther, and Mecerreyes]{Chaix2006}
N.~Chaix, C.~Gourgon, S.~Landis, C.~Perret, M.~Fink, F.~Reuther and
  D.~Mecerreyes, Influence of the molecular weight and imprint conditions on
  the formation of capillary bridges in nanoimprint lithography,
  \emph{Nanotechnology}, 2006, \textbf{17}, 4082\relax
\mciteBstWouldAddEndPuncttrue
\mciteSetBstMidEndSepPunct{\mcitedefaultmidpunct}
{\mcitedefaultendpunct}{\mcitedefaultseppunct}\relax
\EndOfBibitem
\bibitem[Fabi{\`{e}} \emph{et~al.}(2010)Fabi{\`{e}}, Durou, and
  Ondar{\c{c}}uhu]{Fabie2010}
L.~Fabi{\`{e}}, H.~Durou and T.~Ondar{\c{c}}uhu, Capillary {Forces} during
  {Liquid} {Nanodispensing}, \emph{Langmuir}, 2010, \textbf{26},
  1870--1878\relax
\mciteBstWouldAddEndPuncttrue
\mciteSetBstMidEndSepPunct{\mcitedefaultmidpunct}
{\mcitedefaultendpunct}{\mcitedefaultseppunct}\relax
\EndOfBibitem
\bibitem[Eichelsdoerfer \emph{et~al.}(2014)Eichelsdoerfer, Brown, and
  Mirkin]{Eichelsdoerfer2014}
D.~J. Eichelsdoerfer, K.~A. Brown and C.~A. Mirkin, Capillary bridge rupture in
  dip-pen nanolithography, \emph{Soft Matter}, 2014, \textbf{10},
  5603--5608\relax
\mciteBstWouldAddEndPuncttrue
\mciteSetBstMidEndSepPunct{\mcitedefaultmidpunct}
{\mcitedefaultendpunct}{\mcitedefaultseppunct}\relax
\EndOfBibitem
\bibitem[Hunyh \emph{et~al.}(2013)Hunyh, Muradoglu, Liew, and Ng]{Hunyh2013}
T.~Hunyh, M.~Muradoglu, O.~W. Liew and T.~W. Ng, Contact angle and volume
  retention effects from capillary bridge evaporation in biochemical
  microplating, \emph{Colloids and Surfaces A: Physicochemical and Engineering
  Aspects}, 2013, \textbf{436}, 647--655\relax
\mciteBstWouldAddEndPuncttrue
\mciteSetBstMidEndSepPunct{\mcitedefaultmidpunct}
{\mcitedefaultendpunct}{\mcitedefaultseppunct}\relax
\EndOfBibitem
\bibitem[Xu \emph{et~al.}(2006)Xu, Xia, Hong, Lin, Qiu, and Yang]{Xu2006}
J.~Xu, J.~Xia, S.~W. Hong, Z.~Lin, F.~Qiu and Y.~Yang, Self-{Assembly} of
  {Gradient} {Concentric} {Rings} via {Solvent} {Evaporation} from a
  {Capillary} {Bridge}, \emph{Physical Review Letters}, 2006, \textbf{96},
  066104\relax
\mciteBstWouldAddEndPuncttrue
\mciteSetBstMidEndSepPunct{\mcitedefaultmidpunct}
{\mcitedefaultendpunct}{\mcitedefaultseppunct}\relax
\EndOfBibitem
\bibitem[Kumar(2015)]{Kumar2015}
S.~Kumar, Liquid {Transfer} in {Printing} {Processes}: {Liquid} {Bridges} with
  {Moving} {Contact} {Lines}, \emph{Annual Review of Fluid Mechanics}, 2015,
  \textbf{47}, 67--94\relax
\mciteBstWouldAddEndPuncttrue
\mciteSetBstMidEndSepPunct{\mcitedefaultmidpunct}
{\mcitedefaultendpunct}{\mcitedefaultseppunct}\relax
\EndOfBibitem
\bibitem[Anachkov \emph{et~al.}(2016)Anachkov, Lesov, Zanini, Kralchevsky,
  Denkov, and Isa]{Anachkov2016}
S.~E. Anachkov, I.~Lesov, M.~Zanini, P.~A. Kralchevsky, N.~D. Denkov and
  L.~Isa, Particle detachment from fluid interfaces: theory vs. experiments,
  \emph{Soft Matter}, 2016\relax
\mciteBstWouldAddEndPuncttrue
\mciteSetBstMidEndSepPunct{\mcitedefaultmidpunct}
{\mcitedefaultendpunct}{\mcitedefaultseppunct}\relax
\EndOfBibitem
\bibitem[Wu \emph{et~al.}(2016)Wu, Radl, and Khinast]{Wu2016}
M.~Wu, S.~Radl and J.~G. Khinast, A model to predict liquid bridge formation
  between wet particles based on direct numerical simulations, \emph{AIChE
  Journal}, 2016, \textbf{62}, 1877--1897\relax
\mciteBstWouldAddEndPuncttrue
\mciteSetBstMidEndSepPunct{\mcitedefaultmidpunct}
{\mcitedefaultendpunct}{\mcitedefaultseppunct}\relax
\EndOfBibitem
\bibitem[G{\"{o}}gelein \emph{et~al.}(2010)G{\"{o}}gelein, Brinkmann,
  Schr{\"{o}}ter, and Herminghaus]{Gogelein2010}
C.~G{\"{o}}gelein, M.~Brinkmann, M.~Schr{\"{o}}ter and S.~Herminghaus,
  Controlling the {Formation} of {Capillary} {Bridges} in {Binary} {Liquid}
  {Mixtures}, \emph{Langmuir}, 2010, \textbf{26}, 17184--17189\relax
\mciteBstWouldAddEndPuncttrue
\mciteSetBstMidEndSepPunct{\mcitedefaultmidpunct}
{\mcitedefaultendpunct}{\mcitedefaultseppunct}\relax
\EndOfBibitem
\bibitem[Perales and Vega(2011)]{Perales2011}
J.~M. Perales and J.~M. Vega, Dynamics of nearly unstable axisymmetric liquid
  bridges, \emph{Physics of Fluids (1994-present)}, 2011, \textbf{23},
  012107\relax
\mciteBstWouldAddEndPuncttrue
\mciteSetBstMidEndSepPunct{\mcitedefaultmidpunct}
{\mcitedefaultendpunct}{\mcitedefaultseppunct}\relax
\EndOfBibitem
\bibitem[Alexandrou \emph{et~al.}(2010)Alexandrou, Bazilevskii, Entov, Rozhkov,
  and Sharaf]{Alexandrou2010}
A.~N. Alexandrou, A.~V. Bazilevskii, V.~M. Entov, A.~N. Rozhkov and A.~Sharaf,
  Breakup of a capillary bridge of suspensions, \emph{Fluid Dynamics}, 2010,
  \textbf{45}, 952--964\relax
\mciteBstWouldAddEndPuncttrue
\mciteSetBstMidEndSepPunct{\mcitedefaultmidpunct}
{\mcitedefaultendpunct}{\mcitedefaultseppunct}\relax
\EndOfBibitem
\bibitem[Yang \emph{et~al.}(2010)Yang, Tu, and Fang]{Yang2010}
L.~Yang, Y.~Tu and H.~Fang, Modeling the rupture of a capillary liquid bridge
  between a sphere and plane, \emph{Soft Matter}, 2010, \textbf{6},
  6178--6182\relax
\mciteBstWouldAddEndPuncttrue
\mciteSetBstMidEndSepPunct{\mcitedefaultmidpunct}
{\mcitedefaultendpunct}{\mcitedefaultseppunct}\relax
\EndOfBibitem
\bibitem[Men \emph{et~al.}(2011)Men, Zhang, and Wang]{Men2011}
Y.~Men, X.~Zhang and W.~Wang, Rupture kinetics of liquid bridges during a
  pulling process: {A} kinetic density functional theory study, \emph{The
  Journal of Chemical Physics}, 2011, \textbf{134}, 124704\relax
\mciteBstWouldAddEndPuncttrue
\mciteSetBstMidEndSepPunct{\mcitedefaultmidpunct}
{\mcitedefaultendpunct}{\mcitedefaultseppunct}\relax
\EndOfBibitem
\bibitem[Cho \emph{et~al.}(2016)Cho, Hwang, Kim, Lim, Lim, Kim, Gim, and
  Weon]{Cho2016}
K.~Cho, I.~G. Hwang, Y.~Kim, S.~J. Lim, J.~Lim, J.~H. Kim, B.~Gim and B.~M.
  Weon, Low internal pressure in femtoliter water capillary bridges reduces
  evaporation rates, \emph{Scientific Reports}, 2016, \textbf{6}, 22232\relax
\mciteBstWouldAddEndPuncttrue
\mciteSetBstMidEndSepPunct{\mcitedefaultmidpunct}
{\mcitedefaultendpunct}{\mcitedefaultseppunct}\relax
\EndOfBibitem
\bibitem[Neeson \emph{et~al.}(2014)Neeson, Dagastine, Chan, and
  Tabor]{Neeson2014}
M.~J. Neeson, R.~R. Dagastine, D.~Y. Chan and R.~F. Tabor, Evaporation of a
  capillary bridge between a particle and a surface, \emph{Soft Matter}, 2014,
  \textbf{10}, 8489--8499\relax
\mciteBstWouldAddEndPuncttrue
\mciteSetBstMidEndSepPunct{\mcitedefaultmidpunct}
{\mcitedefaultendpunct}{\mcitedefaultseppunct}\relax
\EndOfBibitem
\bibitem[Bayramli \emph{et~al.}(1987)Bayramli, Abou-Obeid, and Van
  De~Ven]{Bayramli1987}
E.~Bayramli, A.~Abou-Obeid and T.~G.~M. Van De~Ven, Liquid bridges between
  spheres in a gravitational field, \emph{Journal of Colloid and Interface
  Science}, 1987, \textbf{116}, 490--502\relax
\mciteBstWouldAddEndPuncttrue
\mciteSetBstMidEndSepPunct{\mcitedefaultmidpunct}
{\mcitedefaultendpunct}{\mcitedefaultseppunct}\relax
\EndOfBibitem
\bibitem[Adams \emph{et~al.}(2002)Adams, Johnson, Seville, and
  Willett]{Adams2002}
M.~J. Adams, S.~A. Johnson, J.~P.~K. Seville and C.~D. Willett, Mapping the
  {Influence} of {Gravity} on {Pendular} {Liquid} {Bridges} between {Rigid}
  {Spheres}, \emph{Langmuir}, 2002, \textbf{18}, 6180--6184\relax
\mciteBstWouldAddEndPuncttrue
\mciteSetBstMidEndSepPunct{\mcitedefaultmidpunct}
{\mcitedefaultendpunct}{\mcitedefaultseppunct}\relax
\EndOfBibitem
\bibitem[Lian and Seville(2016)]{Lian2016}
G.~Lian and J.~Seville, The capillary bridge between two spheres: {New}
  closed-form equations in a two century old problem, \emph{Advances in Colloid
  and Interface Science}, 2016, \textbf{227}, 53--62\relax
\mciteBstWouldAddEndPuncttrue
\mciteSetBstMidEndSepPunct{\mcitedefaultmidpunct}
{\mcitedefaultendpunct}{\mcitedefaultseppunct}\relax
\EndOfBibitem
\bibitem[Mollot \emph{et~al.}(1993)Mollot, Tsamopoulos, Chen, and
  Ashgriz]{Mollot1993}
D.~J. Mollot, J.~Tsamopoulos, T.-Y. Chen and N.~Ashgriz, Nonlinear dynamics of
  capillary bridges : experiments, \emph{Journal of fluid mechanics}, 1993,
  \textbf{255}, 411--435\relax
\mciteBstWouldAddEndPuncttrue
\mciteSetBstMidEndSepPunct{\mcitedefaultmidpunct}
{\mcitedefaultendpunct}{\mcitedefaultseppunct}\relax
\EndOfBibitem
\bibitem[Duprat \emph{et~al.}(2012)Duprat, Proti{\`{e}}re, Beebe, and
  Stone]{Duprat2012}
C.~Duprat, S.~Proti{\`{e}}re, A.~Y. Beebe and H.~A. Stone, Wetting of flexible
  fibre arrays, \emph{Nature}, 2012, \textbf{482}, 510--513\relax
\mciteBstWouldAddEndPuncttrue
\mciteSetBstMidEndSepPunct{\mcitedefaultmidpunct}
{\mcitedefaultendpunct}{\mcitedefaultseppunct}\relax
\EndOfBibitem
\bibitem[Sauret \emph{et~al.}(2015)Sauret, Boulogne, Soh, Dressaire, and
  Stone]{Sauret2015}
A.~Sauret, F.~Boulogne, B.~Soh, E.~Dressaire and H.~A. Stone, Wetting
  morphologies on randomly oriented fibers, \emph{The European Physical Journal
  E}, 2015, \textbf{38}, 62\relax
\mciteBstWouldAddEndPuncttrue
\mciteSetBstMidEndSepPunct{\mcitedefaultmidpunct}
{\mcitedefaultendpunct}{\mcitedefaultseppunct}\relax
\EndOfBibitem
\bibitem[Dejam \emph{et~al.}(2015)Dejam, Hassanzadeh, and Chen]{Dejam2015}
M.~Dejam, H.~Hassanzadeh and Z.~Chen, Capillary forces between two parallel
  plates connected by a liquid bridge, \emph{Journal of Porous Media}, 2015,
  \textbf{18}, 179 --188\relax
\mciteBstWouldAddEndPuncttrue
\mciteSetBstMidEndSepPunct{\mcitedefaultmidpunct}
{\mcitedefaultendpunct}{\mcitedefaultseppunct}\relax
\EndOfBibitem
\bibitem[Cheng and Robbins(2016)]{Cheng2016}
S.~Cheng and M.~O. Robbins, Nanocapillary adhesion between parallel plates,
  \emph{Langmuir}, 2016,  7788--7795\relax
\mciteBstWouldAddEndPuncttrue
\mciteSetBstMidEndSepPunct{\mcitedefaultmidpunct}
{\mcitedefaultendpunct}{\mcitedefaultseppunct}\relax
\EndOfBibitem
\bibitem[Rabinovich \emph{et~al.}(2005)Rabinovich, Esayanur, and
  Moudgil]{Rabinovich2005}
Y.~I. Rabinovich, M.~S. Esayanur and B.~M. Moudgil, Capillary forces between
  two spheres with a fixed volume liquid bridge: theory and experiment,
  \emph{Langmuir}, 2005, \textbf{21}, 10992--10997\relax
\mciteBstWouldAddEndPuncttrue
\mciteSetBstMidEndSepPunct{\mcitedefaultmidpunct}
{\mcitedefaultendpunct}{\mcitedefaultseppunct}\relax
\EndOfBibitem
\bibitem[Dutka and Napi\'orkowski(2007)]{Dutka2007}
F.~Dutka and M.~Napi\'orkowski, The influence of line tension on the formation
  of liquid bridges in atomic force microscope-like geometry, \emph{Journal of
  Physics: Condensed Matter}, 2007, \textbf{19}, 466104\relax
\mciteBstWouldAddEndPuncttrue
\mciteSetBstMidEndSepPunct{\mcitedefaultmidpunct}
{\mcitedefaultendpunct}{\mcitedefaultseppunct}\relax
\EndOfBibitem
\bibitem[Guzowski \emph{et~al.}(2010)Guzowski, Tasinkevych, and
  Dietrich]{Guzowski2010}
J.~Guzowski, M.~Tasinkevych and S.~Dietrich, Free energy of colloidal particles
  at the surface of sessile drops, \emph{The European Physical Journal E},
  2010, \textbf{33}, 219--242\relax
\mciteBstWouldAddEndPuncttrue
\mciteSetBstMidEndSepPunct{\mcitedefaultmidpunct}
{\mcitedefaultendpunct}{\mcitedefaultseppunct}\relax
\EndOfBibitem
\bibitem[D{\"{o}}rmann and Schmid(2014)]{Dormann2014}
M.~D{\"{o}}rmann and H.-J. Schmid, Simulation of capillary bridges between
  nanoscale particles, \emph{Langmuir}, 2014, \textbf{30}, 1055--1062\relax
\mciteBstWouldAddEndPuncttrue
\mciteSetBstMidEndSepPunct{\mcitedefaultmidpunct}
{\mcitedefaultendpunct}{\mcitedefaultseppunct}\relax
\EndOfBibitem
\bibitem[Wang \emph{et~al.}(2016)Wang, Su, Xu, Rong, and Xie]{Wang2016}
L.~Wang, F.~Su, H.~Xu, W.~Rong and H.~Xie, Capillary bridges and capillary
  forces between two axisymmetric power-law particles, \emph{Particuology},
  2016, \textbf{27}, 122--127\relax
\mciteBstWouldAddEndPuncttrue
\mciteSetBstMidEndSepPunct{\mcitedefaultmidpunct}
{\mcitedefaultendpunct}{\mcitedefaultseppunct}\relax
\EndOfBibitem
\bibitem[Dutka and Napi\'orkowski(2006)]{Dutka2006}
F.~Dutka and M.~Napi\'orkowski, Formation of capillary bridges in
  two-dimensional atomic force microscope-like geometry, \emph{The Journal of
  Chemical Physics}, 2006, \textbf{124}, 121101\relax
\mciteBstWouldAddEndPuncttrue
\mciteSetBstMidEndSepPunct{\mcitedefaultmidpunct}
{\mcitedefaultendpunct}{\mcitedefaultseppunct}\relax
\EndOfBibitem
\bibitem[Huh and Mason(1976)]{Huh1976}
C.~Huh and S.~G. Mason, Sphere tensiometry: an evaluation and critique,
  \emph{Can. J. Chem.}, 1976, \textbf{54}, 969--978\relax
\mciteBstWouldAddEndPuncttrue
\mciteSetBstMidEndSepPunct{\mcitedefaultmidpunct}
{\mcitedefaultendpunct}{\mcitedefaultseppunct}\relax
\EndOfBibitem
\bibitem[Bayramli and Mason(1982)]{Bayramli1982}
E.~Bayramli and S.~G. Mason, Some comments on sphere tensiometry, \emph{Colloid
  \& Polymer Sci}, 1982, \textbf{260}, 452--453\relax
\mciteBstWouldAddEndPuncttrue
\mciteSetBstMidEndSepPunct{\mcitedefaultmidpunct}
{\mcitedefaultendpunct}{\mcitedefaultseppunct}\relax
\EndOfBibitem
\bibitem[He \emph{et~al.}(2015)He, Senbil, and Dinsmore]{He2015}
W.~He, N.~Senbil and A.~D. Dinsmore, Measured capillary forces on spheres at
  particle-laden interfaces, \emph{Soft Matter}, 2015, \textbf{11},
  5087--5094\relax
\mciteBstWouldAddEndPuncttrue
\mciteSetBstMidEndSepPunct{\mcitedefaultmidpunct}
{\mcitedefaultendpunct}{\mcitedefaultseppunct}\relax
\EndOfBibitem
\bibitem[Ettelaie and Lishchuk(2015)]{Ettelaie2015}
R.~Ettelaie and S.~V. Lishchuk, Detachment force of particles from fluid
  droplets, \emph{Soft Matter}, 2015, \textbf{11}, 4251--4265\relax
\mciteBstWouldAddEndPuncttrue
\mciteSetBstMidEndSepPunct{\mcitedefaultmidpunct}
{\mcitedefaultendpunct}{\mcitedefaultseppunct}\relax
\EndOfBibitem
\bibitem[Butt \emph{et~al.}(2006)Butt, Graf, and Kappl]{Butt2006}
H.-J. Butt, K.~Graf and M.~Kappl, \emph{Physics and {Chemistry} of
  {Interfaces}}, John Wiley \& Sons, 2006\relax
\mciteBstWouldAddEndPuncttrue
\mciteSetBstMidEndSepPunct{\mcitedefaultmidpunct}
{\mcitedefaultendpunct}{\mcitedefaultseppunct}\relax
\EndOfBibitem
\bibitem[Butt and Kappl(2009)]{Butt2009b}
H.-J. Butt and M.~Kappl, \emph{Surface and {Interfacial} {Forces}}, John Wiley
  \& Sons, 2009\relax
\mciteBstWouldAddEndPuncttrue
\mciteSetBstMidEndSepPunct{\mcitedefaultmidpunct}
{\mcitedefaultendpunct}{\mcitedefaultseppunct}\relax
\EndOfBibitem
\bibitem[J{\o}rgensen \emph{et~al.}(2015)J{\o}rgensen, Merrer,
  Delano{\"{e}}-Ayari, and Barentin]{Jorgensen2015}
L.~J{\o}rgensen, M.~L. Merrer, H.~Delano{\"{e}}-Ayari and C.~Barentin, Yield
  stress and elasticity influence on surface tension measurements, \emph{Soft
  Matter}, 2015, \textbf{11}, 5111--5121\relax
\mciteBstWouldAddEndPuncttrue
\mciteSetBstMidEndSepPunct{\mcitedefaultmidpunct}
{\mcitedefaultendpunct}{\mcitedefaultseppunct}\relax
\EndOfBibitem
\bibitem[Takahashi(1990)]{Takahashi1990}
K.~M. Takahashi, Meniscus shapes on small diameter fibers, \emph{Journal of
  Colloid and Interface Science}, 1990, \textbf{134}, 181--187\relax
\mciteBstWouldAddEndPuncttrue
\mciteSetBstMidEndSepPunct{\mcitedefaultmidpunct}
{\mcitedefaultendpunct}{\mcitedefaultseppunct}\relax
\EndOfBibitem
\bibitem[Qu\'er\'e(1999)]{Quere1999}
D.~Qu\'er\'e, Fluid {Coating} on a {Fiber}, \emph{Annual Review of Fluid
  Mechanics}, 1999, \textbf{31}, 347--384\relax
\mciteBstWouldAddEndPuncttrue
\mciteSetBstMidEndSepPunct{\mcitedefaultmidpunct}
{\mcitedefaultendpunct}{\mcitedefaultseppunct}\relax
\EndOfBibitem
\bibitem[de~Gennes \emph{et~al.}(2004)de~Gennes, Brochard-Wyart, and
  Quere]{Gennes2004}
P.~G. de~Gennes, F.~Brochard-Wyart and D.~Quere, \emph{Capillarity and wetting
  phenomena: drops, bubbles, pearls, waves}, Springer: London, 2004\relax
\mciteBstWouldAddEndPuncttrue
\mciteSetBstMidEndSepPunct{\mcitedefaultmidpunct}
{\mcitedefaultendpunct}{\mcitedefaultseppunct}\relax
\EndOfBibitem
\bibitem[Hubbard(2002)]{Hubbard2002}
A.~T. Hubbard, \emph{Encyclopedia of {Surface} and {Colloid} {Science} -}, CRC
  Press, 2002\relax
\mciteBstWouldAddEndPuncttrue
\mciteSetBstMidEndSepPunct{\mcitedefaultmidpunct}
{\mcitedefaultendpunct}{\mcitedefaultseppunct}\relax
\EndOfBibitem
\bibitem[Drelich \emph{et~al.}(2002)Drelich, Fang, and White]{Drelich2002}
J.~Drelich, C.~Fang and C.~White, Measurement of interfacial tension in
  fluid-fluid systems, \emph{Encyclopedia of surface and colloid science},
  2002, \textbf{3}, 3158--3163\relax
\mciteBstWouldAddEndPuncttrue
\mciteSetBstMidEndSepPunct{\mcitedefaultmidpunct}
{\mcitedefaultendpunct}{\mcitedefaultseppunct}\relax
\EndOfBibitem
\bibitem[Rozynek \emph{et~al.}(2016)Rozynek, Han, Dutka, Garstecki,
  J\'ozefczak, and Luijten]{Rozynek2016}
Z.~Rozynek, M.~Han, F.~Dutka, P.~Garstecki, A.~J\'ozefczak and E.~Luijten,
  \emph{in preparation}, Efficient formation of colloidal chains through
  capillary and dipolar interactions\relax
\mciteBstWouldAddEndPuncttrue
\mciteSetBstMidEndSepPunct{\mcitedefaultmidpunct}
{\mcitedefaultendpunct}{\mcitedefaultseppunct}\relax
\EndOfBibitem
\bibitem[Langbein(2002)]{Langbein2002}
D.~Langbein, \emph{Capillary surfaces: shape - stability - dynamics, in
  particular under weightlessness (Springer Tracts in Modern Physics Vol.178)},
  Springer: Berlin, 2002\relax
\mciteBstWouldAddEndPuncttrue
\mciteSetBstMidEndSepPunct{\mcitedefaultmidpunct}
{\mcitedefaultendpunct}{\mcitedefaultseppunct}\relax
\EndOfBibitem
\bibitem[Boucher(1980)]{Boucher1980}
E.~A. Boucher, Capillary phenomena: {Properties} of systems with fluid/fluid
  interfaces, \emph{Rep. Prog. Phys.}, 1980, \textbf{43}, 497\relax
\mciteBstWouldAddEndPuncttrue
\mciteSetBstMidEndSepPunct{\mcitedefaultmidpunct}
{\mcitedefaultendpunct}{\mcitedefaultseppunct}\relax
\EndOfBibitem
\bibitem[Kralchevsky and Nagayama(2001)]{Kralchevsky2001}
P.~Kralchevsky and K.~Nagayama, \emph{Particles at {Fluid} {Interfaces} and
  {Membranes}: {Attachment} of {Colloid} {Particles} and {Proteins} to
  {Interfaces} and {Formation} of {Two}-{Dimensional} {Arrays}}, Elsevier,
  2001\relax
\mciteBstWouldAddEndPuncttrue
\mciteSetBstMidEndSepPunct{\mcitedefaultmidpunct}
{\mcitedefaultendpunct}{\mcitedefaultseppunct}\relax
\EndOfBibitem
\bibitem[Honschoten \emph{et~al.}(2010)Honschoten, Tas, and
  Elwenspoek]{Honschoten2010}
J.~W.~v. Honschoten, N.~R. Tas and M.~Elwenspoek, The profile of a capillary
  liquid bridge between solid surfaces, \emph{American Journal of Physics},
  2010, \textbf{78}, 277--286\relax
\mciteBstWouldAddEndPuncttrue
\mciteSetBstMidEndSepPunct{\mcitedefaultmidpunct}
{\mcitedefaultendpunct}{\mcitedefaultseppunct}\relax
\EndOfBibitem
\bibitem[Yeomans(1992)]{Yeomans1992}
J.~M. Yeomans, \emph{Statistical mechanics of phase transitions}, Clarendon
  Press, 1992\relax
\mciteBstWouldAddEndPuncttrue
\mciteSetBstMidEndSepPunct{\mcitedefaultmidpunct}
{\mcitedefaultendpunct}{\mcitedefaultseppunct}\relax
\EndOfBibitem
\bibitem[Press(2007)]{Numerical2007}
W.~H. Press, \emph{Numerical {Recipes} 3rd {Edition}: {The} {Art} of
  {Scientific} {Computing}}, Cambridge University Press, 2007\relax
\mciteBstWouldAddEndPuncttrue
\mciteSetBstMidEndSepPunct{\mcitedefaultmidpunct}
{\mcitedefaultendpunct}{\mcitedefaultseppunct}\relax
\EndOfBibitem
\end{mcitethebibliography}
\bibliographystyle{rsc} 

\end{document}